\documentclass[aps,twocolumn,pra,showpacs,floatfix]{revtex4}
\usepackage{epsfig}
\usepackage{dcolumn}

\begin{document}

\title{\bf Highly charged ions for atomic clocks and search for 
variation of the fine structure constant}

\author{V. A. Dzuba and V. V. Flambaum}
\affiliation{School of Physics, University of New South Wales, 
Sydney 2052, Australia}

\date{\today}

\begin{abstract}
We review a number of highly charged ions which have optical transitions suitable for building extremely accurate atomic clocks. This includes ions from Hf$^{12+}$ to U$^{34+}$, which have the $4f^{12}$ configuration of valence electrons, the Ir$^{17+}$ ion, which has a hole in almost filled $4f$ subshell, the Ho$^{14+}$, Cf$^{15+}$, Es$^{17+}$  and Es$^{16+}$ ions. Clock transitions in most of these ions are sensitive to variation of the fine structure constant, $\alpha$ ($\alpha = e^2/\hbar c$). E.g., californium and einsteinium ions have largest known sensitivity to $\alpha$-variation while holmium ion looks as the most suitable ion for experimental study. We study the spectra of the ions and their features relevant to the use as frequency standards. 
\end{abstract} 
\maketitle

Highly charged ions (HCI) can be used for building a new generation of very accurate optical clocks~\cite{HCI,f12}. This may have many technical applications but also clock transitions in HCI can be used to study fundamental problems of modern physics such as variation of fundamental constants~\cite{varHCI}, local Lorentz invariance violation~\cite{LLI1,LLI2}, search for dark matter~\cite{dark}, etc. Having extremely high accuracy of the clocks is crucial for these studies.
Clocks quality factor can be defined as a ratio of clock frequency to the value of its perturbations
\begin{equation}
  Q = \omega/\delta\omega.
\label{e:Q}
\end{equation}
Currenf microwave cesium clock, which serves as definition of metric second, 
have $Q \sim 10^{16}$~\cite{NIST}, best optical clocks approach 
$Q \sim 10^{18}$~\cite{Yb-clock,Sr-clock,Katori} mostly due to larger frequency. Further progress can be achieved by using optical transitions in HCI~\cite{HCI,f12}.
HCI are less sensitive to perturbations due to thier compact size. Therefore,
quality factor $Q=\omega/\delta\omega$ can be larger than in neutral atoms due
to smaller $\delta\omega$. In this paper we review some recent proposals for
very accurate atomic clocks based on HCI and thier use for the search of
time variation of the fine structure constant.

\section{The $4f^{12}$ ions.}

\label{s:f12}

\begin{table}
\caption{\label{t:E2}
Electric quadrupole (E2) clock transition.} 
\begin{ruledtabular}
\begin{tabular}{cccc}
Valence       & Ultra-       & Transition & \\
configuration & relativistic & (ground state - &  $Q$ \\
              & limit        & clock state) & \\
\hline
$p^2$  & $p^2_{1/2}$ & $^3$P$_0 - ^1$D$_2$ or  $^3$P$_0 - ^1$S$_0$ & $<10^{19}$ \\
$p^4$  & $p^2_{3/2}$ & $^3$P$_0 - ^1$S$_0$ & $<10^{19}$ \\
$d^2$  & $d^2_{3/2}$ & $^3$F$_0 - ^3$P$_0$ & $<10^{19}$ \\
$d^8$  & $d^2_{5/2}$ & $^3$F$_4 - ^3$P$_2$ & $ \sim 10^{19}$ \\
$f^2$  & $f^2_{5/2}$ & $^3$H$_4 - ^3$F$_2$ & $ \sim 10^{19}$ \\
$f^{12}$& $f^2_{7/2}$ & $^3$H$_6 - ^3$F$_4$ & $ \sim 10^{20}$ \\
\end{tabular}
\end{ruledtabular}
\end{table}

Optical transitions in HCI can be easily found between states of the same configuration. All such transitions must be even-parity transitions. Even electromagnetic transitions include magnetic dipole (M1), electric quadrupole (E2), and higher-order transitions. Magnetic dipole is usually too strong to be used in clock transitions. Accuracy of the clocks would be limited by natural width of the line. On the other hand, higher-order transitions are too weak and not very conveniet to work with. The best candidates seem to be electric quadrupole transitions (E2). The E2 transitions suitable for the use as clock transitions can be found in configurations consisting of two identical electrons or holes in an almost filled subshell (see Table~\ref{t:E2}).

\begin{figure}
\epsfig{figure=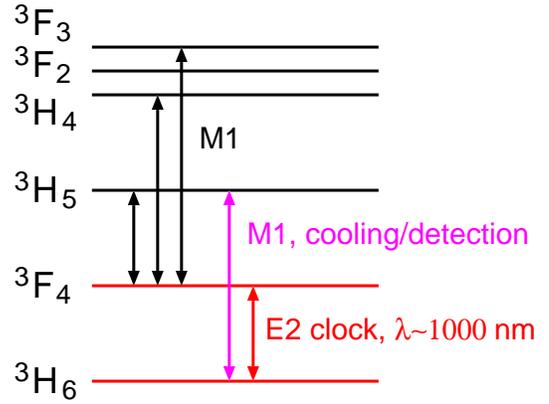,scale=1.0}
\caption{Low-lying energy levels of Os$^{18+}$.}
\label{f:Os18}
\end{figure}

The best candidates seems to be the ions with the $4f^{12}$ configuration of valence electrons in the ground state~\cite{f12}. All HCI from Hf$^{12+}$ to U$^{34+}$ fell in this category. Typical energy diagram is presented on Fig.~\ref{f:Os18} for Os$^{18+}$.

There is an electric quadrupole clock transition between the $^3$H$_6$ and $^3$F$_4$ states as well as magnetic dipole transitions from both ground and clock states, which can be used for cooling and/or detection.

\begin{figure}
\epsfig{figure=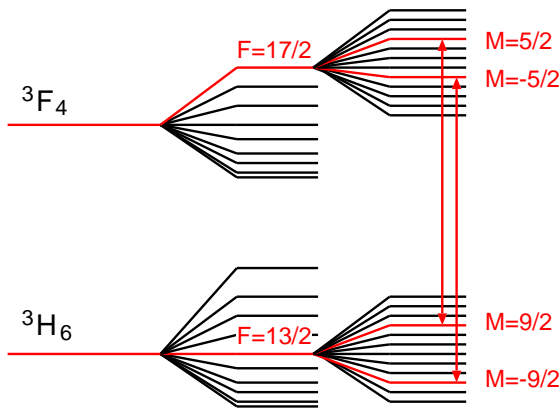,scale=0.8}
\caption{Clock states of Bi$^{25+}$, including hiperfine structure.}
\label{f:Bi25}
\end{figure}

Both states of the clock transition have non-zero quadrupole moments, which make them sensitive to gradients of electric field. This may affect the accuracy of the clock if not addressed. There are different ways to deal with the problem. One is by using the hyperfine structure (hfs). If we take an ion with non-zero nuclear moment, there is a good chance to find hyperfine states with almost identical quadrupole energy shift in upper and lower states~\cite{HCI}. Then the shifts cancell each other in the frequency of the transition (see diagram for $^{209}$Bi$^{25+}$ on Fig.~\ref{f:Bi25}).

This way of dealing with the quadrupole shift has a shorcoming, relatively high sensitivity to the second-order Zeemen shift. This is due to enhancement of the shift by small energy denominators, which are the hfs intervals.

\begin{figure}
\epsfig{figure=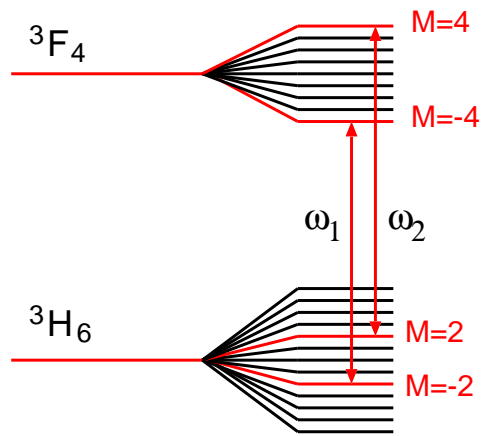,scale=1.0}
\caption{Clock states of Os$^{18+}$, including Zeeman spliting.}
\label{f:f12m}
\end{figure}

Another way of dealing with the problem is by choosing isotopes with zero nuclear spin (and no hyperfine structure)~\cite{f12}. Here one can make a combination of frequencies between different Zeeman states, which is not sensitive to the electric qudrupole shift, see Fig.~\ref{f:f12m}. One has to know the ratio of quadrupole moments of the two clock states to find the right combination~\cite{f12}. Magnetic field is to be used to separate Zeeman states. First order Zeeman shift cancells out in the transition with the same value but opposite sign of the projection $M$ of the total anglular momentum $J$. Second-order Zeeman shift is small.

\section{Search for variation of the fine structure constant.}

The possibility of the fundamental constants to vary is suggested by theories unifying gravity with other fundamental interactions  (see, e.g. \cite{Uzan}).
The study of quasar absorption spectra indicates that the fine structure constant $\alpha$ ($\alpha = e^2/\hbar c$) may vary in space or time. In this study a quasar serves as a powerful source of light in wide range of spectra. Part of this light is absorbed on the way to Earth by a gas cloud brining to Earth informations about atomic spectra billions of years ago. When all the differencies in the quasar absorption spectra and atomic spectra obtained in the laboratory are significantly reduced by adjusting the value of just one parameter, the fine structure constant, this is considered as evidence of slightly different value of $\alpha$ at distant past (or at long distance). The analysis of huge amount of data coming from two telescopes, Kerk telescope in Hawaii and VLT in Chile, reveals that $\alpha$ may vary on astronomical scale along a certain direction in space forming the so called alpha-dipole~\cite{alpha-dipole}
\begin{equation}
  \frac{\Delta \alpha}{\alpha} = \left(1.1 \pm 0.2 \right) \times 10^{-6} \ {\rm Gly}^{-1} \cdot r\cos\theta,
\label{e:adipole}
\end{equation}
where $r$ and $\theta$ are the coordinate of a point in space in the framework of alpha-dipole. 

Earth movements in the framework of alpha-dipole leads to time variation 
of $\alpha$ in laboratory~\cite{BF12}  
\begin{eqnarray}
&&\frac{1}{\alpha}\frac{\partial\alpha}{\partial t} = \left[1.35\times 10^{-18} \cos\psi + \right. \nonumber \\
&&\left. 1.4\times 10^{-20}\cos\omega t \right]{\rm y}^{-1} \approx 10^{-19} {\rm y}^{-1}.
\label{e:alpha-Earth}
\end{eqnarray}
Here $\psi$ is the angle between alpha-dipole and direction of Sun movement, ($\cos\psi \approx 0.07$); second oscillating term in  (\ref{e:alpha-Earth}) is due to Earth movement aroud Sun.

So small rate of change suggests that the most precise atomic clocks are probably the only adequate tools for detecting it. Indeed, the best current limit on the time-variation of $\alpha$ comes from comparing Al$^+$ and Hg$^+$ optical clocks over long period of time~\cite{Al+},
\begin{equation}
\frac{1}{\alpha}\frac{\partial\alpha}{\partial t} = (-1.6 \pm 2.3) \times 10^{-17} {\rm y}^{-1}.
\label{e:Al+}
\end{equation}
The accuracy of the most presice optical clocks approaches the level of $10^{-18}$~\cite{Yb-clock,Sr-clock,Katori}. However, it doen't immediately lead to similar sensitivity to variation of $\alpha$. The relative change in clock frequency due to variation of $\alpha$ can be expressed as 
\begin{equation} 
\frac{1}{\omega}\frac{\partial\omega}{\partial t} = K\frac{1}{\alpha}\frac{\partial\alpha}{\partial t},
\label{e:K}
\end{equation}
where $K$ is electron structure factor which comes from atomic calculations. It turns out that for best optical clocks $K<1$ ($ K=0.31$ for Yb, $K=0.062$ for Sr~~\cite{ADF04}, and $K=-0.3$ for the $f^{12}$ ions considered above). Therefore, we need to search for systems, which have all features of best atomic clocks but also sensitive to variation of $\alpha$, i.e., $K \gg 1$.

It is convenient to present dependence of atomic frequencies on $\alpha$ in a form
\begin{equation}
\omega = \omega_0 + q \left[\left(\alpha/\alpha_0\right)^2-1\right],
\label{e:q}
\end{equation}
where $q$ is the electron stucture factor describing relativistic frequency shift. Comparing (\ref{e:q}) with (\ref{e:K}) leads to $K=2q/\omega_0$. 
There are two ways of searching for large $K$, look for large frequency shift $q$, or look for small frequency $\omega_0$. The best known example of the second kind is dysprosium atom, which has a pair of degenerate states of opposite parity for which $K \sim 10^8$~\cite{Dy}. This pair of states was used indeed in the search for time-variation of the fine structure constant $\alpha$~\cite{Leefer}. The result
\begin{equation}
\frac{1}{\alpha}\frac{\partial\alpha}{\partial t} = (-5.8 \pm 6.9) \times 10^{-17} {\rm y}^{-1}.
\label{e:Dy}
\end{equation}
puts slighly weaker limit on time variation of $\alpha$ than the Al$^+$/Hg$^+$
clocks (\ref{e:Al+}). This is in spite of huge relative enhancement for Dy and almost no enhancement for Al$^+$/Hg$^+$. The reason is that the degenerate states of dysprosium lack the features of an atomic clock transition, e.g. one of the states is pretty short-living. Therefore, what is gained on the enhancement is lost on the accuracy of frequency measurements.

In this work we study another possibility. Keep $\omega_0$ in optical region to take full advantage of extremely accurate optical clocks, and find systems with large relativistic energy shift $q$. To see where such systems can be found, it is instructive to use an analitical estimate for relativistic energy shift~\cite{DFW99}
\begin{equation}
\Delta E \approx \frac{E}{\nu} \left(Z\alpha\right)^2\left(\frac{1}{j+1/2}-C\right),
\label{e:DE}
\end{equation}
where $\nu$ is the effective principal quantum number ($E=-1/2\nu^2$), $Z$ is nuclear charge, $j$ is total angular electron momentum, and $C$ is semi-empirical factor to simulate many-body effects in many-electron atoms ($C \sim 0.6$). Relativistic frequency shift for a transition between states $a$ and $b$ is given by $q= \Delta E_a - \Delta E_b$. One can see from (\ref{e:DE}) that large frequency shift $q$ can be found in heavy (large $Z$) highly charged ions, where $E \sim (Z_i+1)^2$ ($Z_i$ is ionization degree), in transitions, which correspond to $s - f$ or $p - f$ single-electron transitions (largest change of $j$)~\cite{varHCI}. 
One problem here is that such transitions are usually not optical since energy intervals grow very fast with ionization degree $Z_i$ ($\Delta E \sim (Z_i+1)^2$). The solution comes from level crossing~\cite{varHCI,crossing,hole}. Since state ordering in neutral atoms and hydrogen-like ions is different, there are must be change in ordering of $s$ and $f$ or $p$ and $f$ states at some intermediate ionization degree. In the vicitity of level crossing the frequencies of corresponding transitions are likely to be in optical region~\cite{crossing}. 

A number of optical transitions in HCI sensitive to variation of the fine structure constant were considered in Ref.~\cite{varHCI,hole,one-el,Ag-like,Cd-like}. However, most of these transitions lack some features of atomic clock transitions, which limit the accuracy of frequency measurements. Important questions of ions traping and cooling, preparation and detection of states, etc. were also not discussed. All these questions were first addressed in recent paper~\cite{Ho14}. In particular, the criteria for good clock system, sensitive to variation of $\alpha$, were formulated. The main points include: (a) In terms of single-electron transitions, the clock transition is a $s - f$ or $p - f$ transition. This makes it sensitive to the variation of $\alpha$. (b) the frequency of the transition is in optical region ($5000 \ {\rm cm}^{-1} < \hbar\omega < 43000 \ {\rm cm}^{-1}$, or $230 \ {\rm nm} < \lambda < 2000 \ {\rm nm}$). (c) This is a transition bewteen ground and a metastable state with lifetime between 100 and $10^4$ seconds. (d) There are other relatively strong optical transitions with transition rate $\gtrsim 10^3 \ {\rm s}^{-1}$. (e) The transition is not sensitive to perturbations such as gradients of electric field, Zeeman shift, black-body radiation shift, etc.
 
A very promising systems is the Ho$^{14+}$ ion~\cite{Ho14}. 
It has following features: (a) Clock transition between the $4f^65s \ ^8{\rm F}_{1/2}$ and $4f^55s^2 \ ^6{\rm H}^o_{5/2}$ states is sensitive to variation of alpha (the $4f - 5s$ transition). (b) It is optical transition ($\lambda \approx 400$~nm). (c) It is a narrow transition from ground state to a metastable state. (d) There are electric dipole (E1) and magnetic dipole (M1) transitions from both ground and clock states. (e) The clock transition can be made insensitive to gradients of electric field, which are coupled to atomic quadrupole moment. This can be done by proper choice of the values of the total angular momentum $F$ (including nuclear spin $I$, $F=J+I$) and its projection $M$. Quadrupole shift disappears for $F=3$ and 
$M=2$ since it is proportional to $3M^2-F(F+1)$. 
Experimental work with the Ho$^{14+}$ ions is in progress at RIKEN~\cite{RIKEN}.

There must be at least two clocks to register variation of alpha since such variation can only be unambiguously detected in a dimensionless ratio of two
frequencies. A good option is to have one clock transition, which is sensitive to variation of $\alpha$ and another, which is not. Note that cesium clock is not good enough if we want relative accuracy of the order $10^{-18}$. A good option might come from the use of HCI with the $4f^{12}$ configuration of
valence electrons considered in section \ref{s:f12}, e.g. the Os$^{18+}$ ion.
Comparing the Ho$^{14+}$ and Os$^{18+}$ clocks provides high
sensitivity to variation of alpha ($K \approx -18$).

\begin{figure}
\epsfig{figure=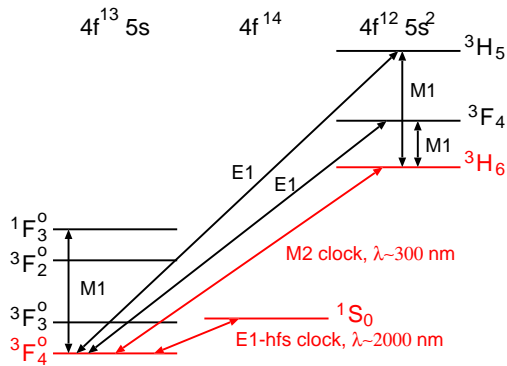,scale=0.65}
\caption{Low-lying energy levels of Ir$^{17+}$.}
\label{f:Ir17}
\end{figure}

\begin{table}
\caption{\label{t:ir17}
Excitation energies ($E$, cm$^{-1}$), sensitivity factors ($q$, cm$^{-1}$), and enhansement factors ($K = 2q/E$) for the clock states of Ir$^{17+}$ ion.}
\begin{ruledtabular}
\begin{tabular}{rlc rrr}
\multicolumn{1}{c}{N}& \multicolumn{1}{c}{Conf.}&
\multicolumn{1}{c}{Term}&
\multicolumn{1}{c}{$E$}&
\multicolumn{1}{c}{$q$}&
\multicolumn{1}{c}{$K$}\\
\hline
1 &$4f^{13}5s$  &$^2$F$^o_{4}$&     0 & 0 & 0 \\
2 &$4f^{14}$    &$^1$S$_{0}$&  5000 & 370000 & 150 \\
3 &$4f^{12}5s^2$&$^3$H$_{6}$& 30000 & -390000 & -26 \\
\end{tabular}
\end{ruledtabular}
\end{table}

\begin{table*}
\caption{\label{t:CfEs}
Long-living isotopes of Cf and Es and clock transitions in Cf$^{15+}$, Es$^{17+}$ and Es$^{16+}$.} 
\begin{ruledtabular}
\begin{tabular}{lcc llll rrr}
\multicolumn{3}{c}{Isotope}& 
\multicolumn{4}{c}{Clock transition}&
\multicolumn{1}{c}{$\hbar\omega$}&
\multicolumn{1}{c}{$q$}&
\multicolumn{1}{c}{$K$}\\
\multicolumn{1}{c}{Ion}&
\multicolumn{1}{c}{Lifetime}&
\multicolumn{1}{c}{$I$}&
\multicolumn{2}{c}{Ground state}&
\multicolumn{2}{c}{Clock state}&
\multicolumn{1}{c}{cm$^{-1}$}&
\multicolumn{1}{c}{cm$^{-1}$}& \\
\hline
$^{249}$Cf$^{15+}$ & 351 y & 9/2 & $5f6p^2$ & $ ^2$F$^o_{5/2}$ & $5f^26p$ & $ ^2$H$^o_{9/2}$ & 13303 & 380000 & 50 \\

$^{252}$Es$^{17+}$ & 1.29 y & 5 & $5f^2$ & $^3$H$_{4}$ & $5f6p$ & $^3$F$_{2}$ & 7017 & -46600 & -13 \\

$^{253}$Es$^{16+}$ & 20 d & 7/2 & $5f6p^2$ & $ ^2$F$^o_{5/2}$ & $5f^26p$ & $ ^2$H$^o_{9/2}$ & 7475 & -184000 & -63 \\

\end{tabular}
\end{ruledtabular}
\end{table*}

Another option is to use the Ir$^{17+}$ ions. Clock transitions in these ions involve hole states in the $4f$ subshell which leads to extra enhancement of the sensitivity of clock frequencies to variation of the fine structure constant~\cite{hole}. A diagram for few lowest states of Ir$^{17+}$ is presented on Fig.~\ref{f:Ir17}. The energies and sensitivety coefficients for clock states of Ir$^{17+}$ ions are presented in Table~\ref{t:ir17}. Note, that the ion has two clock transitions which have different sensitivity to variation of $\alpha$. If the ratio of two frequencies is measured, the combined sensitivity is very large
\begin{equation}
\frac{\partial}{\partial t}\ln\frac{\omega_2}{\omega_1} = \left(K_2 - K_1\right)\frac{1}{\alpha}\frac{\partial\alpha}{\partial t} = -176\frac{1}{\alpha}\frac{\partial\alpha}{\partial t}.
\label{e:ir17}
\end{equation}
Experimental work with Ir$^{17+}$ ions is in progress at Max Planck Institute~\cite{Ir17}. 

It has been mentioned above that we are looking for systems with large sensitivity coefficients $q$ and clock frequency $\omega$ beeing in optical region so that the enhancement factor $K$ ($K=2q/\omega$) is large due to large $q$ rather than small $\omega$. The largest sensitivity factors found so far are in the Cf$^{17+}$ and Cf$^{16+}$ ions~\cite{Cf}. However, both these ions are not very convenient for the use as atomic clocks. The  Cf$^{17+}$ ion has only one optical transition other than clock transition. This is a magnetic dipole transition between ground $5f_{5/2}$ state and excited $5f_{7/2}$ state. It is weak due to small frequency of the transition. This makes it difficult working with the ion. The  Cf$^{16+}$ ion has only one metastable excited state which can serve as a clock state. The transition between ground $5f6p \ ^1{\rm F}_3$ and metastable $6p^2 \ ^1{\rm S}_0$ states is the magnetic octupole (M3) transition. It is so weak that its use as clock transition is very problematic. 

It turns out that good clock transitions exist in the Cf$^{15+}$ ion as well as in the Es$^{16+}$ and Es$^{17+}$ ions. The main parameters of the ions and corresponding clock transitions are summarized in Table~\ref{t:CfEs}. Note that both elements have only unstable isotopes. However, many isotopes have very long lifetime (see, e.g. Table~\ref{t:CfEs}). The choice of isotopes in Table~\ref{t:CfEs} is dictated by two considerations. First, it is a long-living isotope. Second, nuclear spin $I$ has such a value that it is always possible to have $F=3$, $M=2$ for both states in the clock transition. Here $F=I+J$ is the total angular momentiom of the ion and $M$ is its projection. States with $F=3$ and $M=2$ have zero quadrupole moment and are not sensitive to gradients of electric field.

The authors are gratefull to H. Katori and M. Wada for useful discussions.
The work was supported by the Australian Research Council.


%

\end{document}